\documentclass{article}
\usepackage{ismir,amsmath,cite,url}
\usepackage{graphicx}
\usepackage{color}

\usepackage{cite}
\usepackage{amsmath,amssymb,amsfonts}
\usepackage{graphicx}
\usepackage{textcomp}

\usepackage{booktabs}
\usepackage{multirow}
\usepackage{nicefrac}
\usepackage[caption=false]{subfig}
\usepackage[inline]{enumitem}
\usepackage{url}
\usepackage[group-four-digits]{siunitx}
\usepackage[colorinlistoftodos,prependcaption,textsize=tiny]{todonotes}
\usepackage{microtype}

\title{Genre-Agnostic Key Classification \\ With Convolutional Neural Networks}

\oneauthor
{Filip Korzeniowski and Gerhard Widmer}
{Institute of Computational Perception,\\
Johannes Kepler University, Linz, Austria\\
filip.korzeniowski@jku.at}

\newcommand{\keynet}{KeyNet}
\newcommand{\keynetfull}{KeyNet/F}
\newcommand{\keynetsnip}{KeyNet/S}
\newcommand{\allconv}{AllConv}

\begin{document}

\maketitle
\begin{abstract}
We propose modifications to the model structure and training procedure to a
recently introduced Convolutional Neural Network for musical key
classification. These modifications enable the network to learn a
genre-independent model that performs better than models trained for specific
music styles, which has not been the case in existing work. We analyse this
generalisation capability on three datasets comprising distinct genres. We
then evaluate the model on a number of unseen data sets, and show its superior
performance compared to the state of the art. Finally, we investigate the
model's performance on short excerpts of audio. From these experiments, we
conclude that models need to consider the harmonic coherence of the whole piece
when classifying the local key of short segments of audio.
\end{abstract}
\section{Introduction}

The musical key is the highest-level harmonic representation in Western tonal
music. It thus plays a central role in understanding the semantic content of a
piece. Such understanding drives not only theoretical analyses of music, but is
also relevant for modern music creators, who mix samples from various different
pieces that fit well harmonically into a new composition. However, deriving the
key of a musical piece is a demanding task that only experts can perform. It
is thus impractical to annotate large music collections by hand. Therefore,
we need computational key classification systems.

Most key classification systems
(e.g.~\cite{pauws_musical_2004,temperley_what_1999,noland_signal_2007,faraldo_key_2016})
conform to the same principle: they extract a time-frequency representation of
the audio, filter out nuisances, map this representation to chroma vectors, and
accumulate them over time. The resulting feature vector is then compared to
template vectors for each key. The drawbacks of such approaches include that
key templates differ for different musical genres~\cite{faraldo_key_2016} and
favour one key mode over another~\cite{albrecht_use_2013}. This leads to key
classification systems that perform well only on the musical styles they were
designed for.  Although there are attempts to address these
issues~\cite{bernardes_automatic_2017}, ideally, we would want a model that
handles different kinds of input autonomously, and does not need human intervention to
e.g.\ balance mode probabilities.

Data-driven methods bear the potential to meet this requirement. Recently, an
end-to-end neural-network-based key classification model was
introduced~\cite{korzeniowski_endtoend_2017}. Although it generalised better
across musical genres than hand-crafted approaches, it still achieved the best
results when tuned specifically for a musical style.  In this paper, we present
modifications to the model structure and its training procedure that enable the
model to learn a key classifier that is agnostic to genre. Not only does it
perform better than the model proposed in~\cite{korzeniowski_endtoend_2017} on
all genres the latter is optimised for; it does so not despite, but
\emph{because} it is trained on various musical styles, instead of a specific
one (see Sec.~\ref{sec:influence_of_training_data}).

\section{Method}

We build upon the same audio processing pipeline used
in~\cite{korzeniowski_endtoend_2017}, and input to the network a log-magnitude
log-frequency spectrogram (5 frames per second, frame size
\num{8192}, sample rate \SI{44100}{\Hz}, 24 bins per octave). We limit the
frequency range to the harmonically most relevant \SI{65}{\Hz} to
\SI{2100}{\Hz}, as found in~\cite{korzeniowski_feature_2016}.

The network structure proposed in~\cite{korzeniowski_endtoend_2017} was
modelled after typical processing pipelines used for key classification. It
features five convolutional layers of $5\times5$ kernels for spectrogram
processing, followed by a dense projection into a frame-wise embedding space,
which is then averaged over time and classified using a softmax layer. All
layers except the last use the exponential-linear activation
function~\cite{clevert_fast_2016} (ELU). The architecture, which we name
\emph{\keynet{}}, is summarised in Table~\ref{tab:keynet_arch}.

During training, the model is shown the complete spectrogram of a piece. Its
weights are then adapted using stochastic gradient descent to minimise the
categorical cross-entropy between the predicted key distribution and the ground
truth. We will refer to the \keynet{} architecture, when trained using full spectrograms, as \emph{\keynetfull}.

\subsection{Adaptations of the Training Procedure}

The outlined training scheme has two drawbacks. First, the computation of a
single update is expensive; the network has to process the full spectrogram
(e.g.\ \num{600}$\times$\num{105} values for a two-minute piece), and keep
intermediate results for back-propagating the error. Training is thus slow and
requires much memory. Second, it keeps the variety of the data lower than
necessary, as the network sees the same spectrograms at every epoch.

To circumvent these drawbacks, we show the network only short snippets instead
of the whole piece at training time (similar to random cropping in computer
vision). These snippets should be as short as possible to reduce computation
time, but have to be long enough to contain the relevant information to
determine the key of a piece.  From our datasets, we found \SI{20}{\second} to
be sufficient (with the exception of classical music, which we need to treat
differently, due to the possibility of extended periods of modulation---see
Sec.~\ref{sec:data} below).
Each time the network is presented a song, we cut a random \SI{20}{\second}
snippet from the spectrogram. The network thus sees a different variation of
each song every epoch. 

During testing, the network processes the whole piece. This gives better
results than when using only a snippet. Since we do not have to store
intermediate results and process each piece many times as in training,
memory space and run time are not an issue. We will refer to \keynet{} models
trained using spectrogram snippets as \emph{\keynetsnip{}}.

We expect this modification to have the following effects.
\begin{enumerate*}[label=\alph*)]
\item Back-propagation will be faster and require less memory, because the network sees shorter snippets; we can thus train faster, and process larger models.
\item The network will be less prone to over-fitting, since it almost never sees the same training input; we expect the model to generalise better.
\item The network will be forced to find evidence for a key in each excerpt of the training pieces, instead of relying on parts where the key is more obvious; by asking more of the model, we expect it to pick up more subtle relationships between the audio and its key.
\end{enumerate*}

\subsection{Adaptations of the Model Structure}

The \keynet{} architecture uses a dense layer to project the processed
spectrogram into a key embedding space. In its original formulation, which uses
an embedding space with 48 dimensions and 8 feature maps in the convolutional
layers, this projection accounts for \SI{65}{\percent} of the network's
parameters. Dense layers are also more prone to over-fitting than convolutional
layers.

We thus propose to use a network architecture that does away with dense layers,
and relies on convolutions and pooling only. At the same time, we move away
from modelling the network based on traditional key classification methods---recall that the components of \keynet{} were designed to correspond to components in typical key classification pipelines---and instead use a general
network architecture for classification, based on the all-convolutional net~\cite{springenberg_striving_2014}. The
new architecture is summarised in Table~\ref{tab:allconv_arch}, and will be
referred to as \emph{\allconv{}}. As with \keynetsnip{}, we will train this
architecture only with the snippet method.

We expect this change to improve results and generalisation because
\begin{enumerate*}[label=\alph*)]
\item convolutional layers over-fit less than dense layers;
\item given the same number of parameters, deeper networks are more expressive than shallower ones~\cite{eldan_power_2016,liang_why_2017};
\item comparable architectures have shown to perform well in other audio-related tasks~\cite{eghbal-zadeh_cpjku_2016,korzeniowski_fully_2016}.
\end{enumerate*}

\newcommand{\nf}{N_f}

\begin{table}[t!]
\centering
\subfloat[\keynet{} Architecture]{\footnotesize
\begin{tabular}{@{}lrr@{}}
\toprule
\textbf{Layer Type} & \textbf{FMaps} & \textbf{Params} \\ \midrule
Input               &                &                 \\
Conv-ELU            & $\nf$          & $5\times5$      \\ \midrule
Conv-ELU            & $\nf$          & $5\times5$      \\ \midrule
Conv-ELU            & $\nf$          & $5\times5$      \\ \midrule
Conv-ELU            & $\nf$          & $5\times5$      \\ \midrule
Conv-ELU            & $\nf$          & $5\times5$      \\ \midrule
Dense-ELU           &                & $2\cdot\nf$ \\
\multicolumn{2}{@{}l}{Pool-Time Avg.} & \\
\multicolumn{2}{@{}l}{Dense-Softmax}                  & $24$ \\ \bottomrule
\end{tabular}\label{tab:keynet_arch}}
\hfill
\subfloat[\allconv{} Architecture]{\footnotesize
\begin{tabular}{@{}lrr@{}}
\toprule
\textbf{Layer Type} & \textbf{FMaps} & \textbf{Params} \\ \midrule
Input               &                &                 \\
Conv-ELU            & $\nf$          & $5\times5$      \\
Conv-ELU            & $\nf$          & $3\times3$      \\
Pool-Max            &                & $2\times2$      \\ \midrule
Conv-ELU            & $2\nf$         & $3\times3$      \\
Conv-ELU            & $2\nf$         & $3\times3$      \\
Pool-Max            &                & $2\times2$      \\ \midrule
Conv-ELU            & $4\nf$         & $3\times3$      \\
Conv-ELU            & $4\nf$         & $3\times3$      \\
Pool-Max            &                & $2\times2$      \\ \midrule
Conv-ELU            & $8\nf$         & $3\times3$      \\ \midrule
Conv-ELU            & $8\nf$         & $3\times3$      \\ \midrule
Conv-ELU            & $24$           & $1\times1$      \\
\multicolumn{2}{@{}l}{Pool-Global Avg.} &                 \\
Softmax             &                &                 \\ \bottomrule
\end{tabular}\label{tab:allconv_arch}}
\caption{Neural Network architectures. $\nf$ is a parameter that controls the
model complexity. Horizontal lines denote dropout
layers~\cite{srivastava_dropout_2014}. Here, dropout is
applied on complete feature maps, not individual units. Each convolution is
followed by batch normalisation~\cite{ioffe_batch_2015}. \emph{FMaps} indicates
the number of feature maps, while \emph{Params} the parameters of the layer
(kernel size, pool size, or number of units).}
\label{tab:architectures}
\end{table}

\section{Experiments}

We first evaluate how the proposed modifications affect the key
classification performance in Sec.~\ref{sec:evaluation_adaptations}. Then,
we analyse how the number and genre of training data influence results in
Sec.~\ref{sec:influence_of_training_data}.

\subsection{Data}
\label{sec:data}

Since we are interested in how well the models generalise across different
genres, we use datasets that encompass three distinct musical styles. As
in~\cite{korzeniowski_endtoend_2017}, we apply pitch shifting in the range of
-4 to +7 semitones to increase the amount of training data.

\begin{description}
\item[Electronic Dance Music:]
Here, we use songs from the GiantSteps MTG Key
dataset\footnote{\url{https://github.com/GiantSteps/giantsteps-mtg-key-dataset}},
collected by \'Angel Faraldo. It comprises 1486 distinct two-minute audio
previews from \url{www.beatport.com}, with key ground truth for each excerpt. We only use excerpts labelled with a single key and a high confidence (1077 pieces), and split them into \SI{80}{\percent} training and \SI{20}{\percent} validation.
For testing, we use the GiantSteps Key Dataset\footnote{\url{https://github.com/GiantSteps/giantsteps-key-dataset}}.
It comprises 604 two-minute audio previews from the same source (but distinct from the training set).
\pagebreak
\item[Pop/Rock Music:]
For this genre, we use the McGill Billboard
dataset~\cite{burgoyne_expert_2011}\footnote{\url{http://ddmal.music.mcgill.ca/research/billboard}}.
It consists of 742 unique songs sampled from the American Billboard charts
between 1958 and 1991. We split these songs into subsets of 62.5\% for
training, 12.5\% for validation, and 25\% for testing. We determine the global
key for each song using the procedure described
in~\cite{korzeniowski_endtoend_2017}, which leaves us with 625 songs with key
annotations in total. The exact division and key ground truths are available
online\footnote{\url{http://www.cp.jku.at/people/korzeniowski/bb.zip}}.
\item[Classical Music:]
To cover this genre, we collected 1504 (mostly piano) pieces from our internal
database for which we could derive the key from the piece's title. Classical
pieces often modulate their key, but usually start in the key denoted in the
title. We thus only use the first \SI{30}{\second} of each recording. Tracking
key modulations is left for future work. We then select \SI{81}{\percent} for
training, \SI{9}{\percent} for validation, and \SI{10}{\percent} for testing.
\end{description}

\subsection{Metrics}

We adopt the standard evaluation score for Key Classification as defined in the
MIREX evaluation campaign\footnote{\url{http://www.music-ir.org/mirex}}. It
goes beyond simple accuracy, as it considers harmonic similarities between
key classes. A prediction can fall into one of the following categories:
\begin{description}
\item[Correct:] if the tonic and the mode (major/minor) of
        prediction and target correspond.
\item[Fifth:] if the tonic of the prediction is the fifth of
        the target (or vice versa), and modes correspond.
\item[Relative Minor/Major:] if modes differ and either
        \begin{enumerate*}[label=\alph*)]
        \item the predicted mode is minor and the predicted tonic is 3
        semitones below the target, or
        \item the predicted mode is major and the predicted tonic is 3
        semitones above the target.
        \end{enumerate*}
\item[Parallel Minor/Major:] if modes differ but the predicted
        tonic matches the target.
\item[Other:] Prediction errors not caught by any category,
        i.e.\ the most severe errors.
\end{description}
Then, a weighted score can be computed as $w = r_c + 0.5 \cdot r_f + 0.3
\cdot r_r + 0.2 \cdot r_p$, where $r_c$, $r_f$, $r_r$, and $r_p$ are the ratios
of the correct, fifth, relative minor/major, and parallel minor/major,
respectively. We will use this weighted score for our comparisons.

\subsection{Evaluation of the Adaptations}
\label{sec:evaluation_adaptations}

\begin{figure}[t]
\centering
\includegraphics{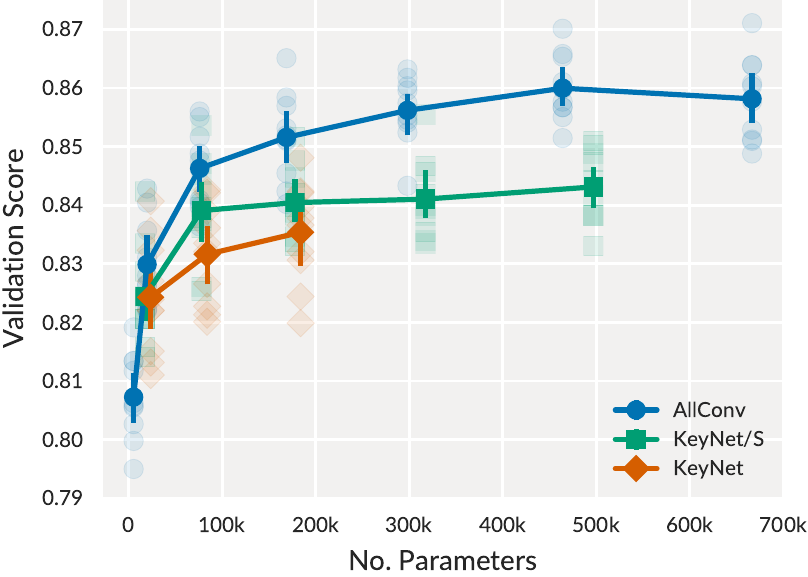}
\caption{Average validation score over 10 runs for the different model setups.
         Whiskers represent \SI{95}{\percent} confidence intervals computed by
         bootstrapping. Transparent dots show results of the individual runs.
         We see that given similar network sizes, the \allconv{} model performs best.
         Also, using snippet training (\keynetsnip{}) improves results compared to
         full spectrogram training (\keynetfull{}), and enables training larger
         networks.}
\label{fig:gridsearch}
\vspace{-.15cm}
\end{figure}

To evaluate the effect of our proposed adaptations, we train the three setups
(\keynetfull{}, \keynetsnip{}, \allconv{}) with the combined data of all datasets. We will
consider \emph{validation} results in the first sets of experiments, and show
results on the testing sets only for our analyses and final evaluations. This
way, we ensure that the final results are unbiased.

The capacity of a neural network depends not only on the architecture,
but also on its size. For a fair comparison,
we evaluate each architecture with varying network sizes. For the \allconv{}
architecture, we select the number of feature maps $\nf \in \{2, 4, 8, 12, 16,
20, 24\}$. For the \keynet{} architecture, the network size depends on the
number of feature maps in the convolutional layers and the size of the
embedding space. For practical reasons, we set the size of the embedding space
to be $2\nf$, and select $\nf \in \{8, 16, 24, 32, 40\}$. Note that if we train
on full spectrograms (\keynetfull{}), we could not train networks with $\nf >
24$ due to memory constraints. For each model, we tried dropout probabilities
of $p \in \{0.0,0.1, 0.2\}$.

Figure~\ref{fig:gridsearch} presents the results of the three model
configurations. For each model and model capacity, we select the best dropout
probability based on the validation results. The experiments show that both
adaptations are beneficial. Training with snippets instead of full spectrograms gives better results at smaller network capacities and enables training of larger networks. The \allconv{} architecture achieves even better results, regardless of its size.

We can quantify two reasons for this, which are consequences of the expected benefits of the adaptations: \emph{better generalization} through increased data variety and the absence of dense layers, and \emph{better expressivity} through deeper architectures and by training the network on a more difficult task. For the first, better generalisation, we compare the average ratio of validation accuracy to training accuracy for each of the models (higher indicates less over-fitting): \num{0.944852}, \num{0.968913}, and \num{0.981850} for \keynetfull{}, \keynetsnip{}, and \allconv{}, respectively. For the second, \emph{model expressiveness}, we compare the model's capability to fit the training data in terms of accuracy: \num{0.836884261113}, \num{0.858029635563}, and \num{0.906928313977} for \keynetfull{}, \keynetsnip{}, and \allconv{}, respectively. Stronger models that generalise better achieve better results.

\subsection{Influence of Training Data}
\label{sec:influence_of_training_data}

\begin{figure}[t!]
\centering
\includegraphics{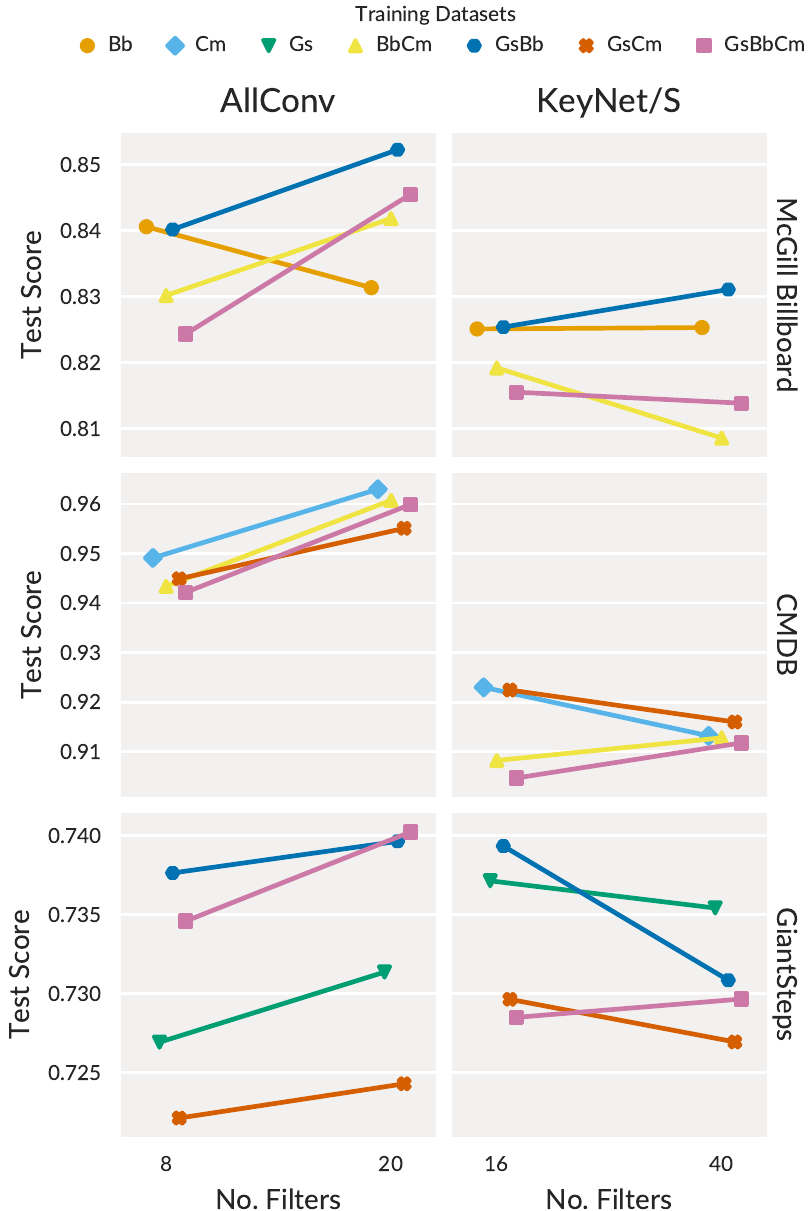}
\caption{Average test scores over 10 runs for each architecture (columns),
        split by dataset (rows).  The smaller models are on the left of each
        column. Colors indicate the training data used: \emph{Bb} stands for
        the Billboard dataset, \emph{Cm} for the classical music dataset, and
        \emph{Gs} for the GiantSteps dataset. Each row shows the results of runs where the training set also contained the training data of the respective test set genre (e.g.\ in the first row, we
        only see runs where McGill Billboard data was included in training).}
\label{fig:dataset_influences}
\end{figure}

We then want to see how the number and genre of the datasets used for training affects
results. To this end, we select the hyper-parameter settings for \allconv{} and
\keynetsnip{} that achieved the best average results in the previous experiment:
$\nf=20, p=0.1$ for \allconv{}, $\nf=40, p=0.1$ for \keynetsnip{}. Additionally, we
consider smaller models of each type, i.e. $\nf=8$ for \allconv{} and $\nf=16$ for
\keynetsnip{}, both without dropout. Under these settings, both architectures have
a comparable number of parameters. We train these models using all possible 1,
2, and 3-combinations of the datasets, and evaluate them on all data. The
results are shown in Fig.~\ref{fig:dataset_influences}.

The main observations are:
\begin{enumerate*}[label=\alph*)]
\item increasing model capacity is more beneficial to the \allconv{} model than
      \keynetsnip{}, regardless of dataset;
\item adding capacity to the \allconv{} model enables it to better deal with
      diverse data---the biggest gains of additional parameters are achieved
      if the model is trained on a combined dataset (pink line)---while this
      is not always the case for \keynetsnip{} (see the Billboard results, where
      it seems that adding classical music to the training set impairs the
      performance of this model);
\item given enough capacity in the \allconv{} model, training using the complete
      data performs better than (or almost equal to) fitting a specific genre,
      while the opposite is the case for \keynetsnip{}, where specialised models
      outperform the general ones.
\end{enumerate*}
We thus argue that the \allconv{} model not only copes better with diverse training data, but that it leverages the diversity in the training data to perform as well as it does.

\begin{table*}[t!]
\centering
\newcommand{\myspace}{0.8em}
\begin{tabular}{@{}llrrrrrr@{}}
\toprule
\textbf{Dataset} & \textbf{Model}                                          & \textbf{Weighted}    & \textbf{Correct}    & \textbf{Fifth} & \textbf{Relative} & \textbf{Parallel} & \textbf{Other}    \\ \midrule
GiantSteps       & AllConv                                                 & \textbf{74.6} & \textbf{67.9} & 7.0        & 8.1        & 4.1        & \textbf{12.9} \\
                 & CK1~\cite{korzeniowski_endtoend_2017} & 74.3          & \textbf{67.9} & 6.8        & 7.1        & 4.3        & 13.9          \\[\myspace]
Billboard        & AllConv                                                 & \textbf{85.1} & \textbf{79.9} & 5.6        & 4.2        & 6.2        & \textbf{4.2}  \\
                 & CK2~\cite{korzeniowski_endtoend_2017} & 83.9          & 77.1          & 9.0        & 4.9        & 4.2        & 4.9           \\[\myspace]
Classical  & AllConv                                                 & 96.6          & 95.2          & 1.4        & 1.4        & 1.4        & 0.7           \\
                 & -                                                       & -             & -             & -          & -          & -          & -             \\[\myspace]
KeyFinder        & AllConv                                                 & \textbf{76.1} & \textbf{70.0} & 5.7        & 7.4        & 4.7        & \textbf{12.1} \\
                 & bgate \cite{faraldo_multiprofile_2017}                  & 72.4          & 65.0          & 8.6        & 6.5        & 5.4        & 14.4          \\[\myspace]
Isophonics       & AllConv                                                 & \textbf{82.5} & \textbf{76.3} & 7.6        & 5.4        & 3.7        & \textbf{7.1}  \\
                 & BD1 \cite{bernardes_automatic_2017}                     & 75.1          & 66.0          & 13.6       & 5.1        & 3.9        & 9.2           \\[\myspace]
R. Williams  & AllConv                                                 & \textbf{81.2} & \textbf{72.4} & 10.8       & 10.3       & 1.3        & \textbf{5.2}  \\
                 & HS1 \cite{schreiber_mirex_2017}                         & 77.1          & 68.8          & 10.1       & 9.0        & 3.2        & 9.0           \\[\myspace]
Rock             & AllConv                                                 & 74.3          & 69.3          & 6.5        & 1.7        & 6.0        & 16.5          \\
                 & -                                                       & -             & -             & -          & -          & -          & -             \\ \bottomrule
\end{tabular}
\caption{Evaluation results. Best results are in boldface.}
\label{tab:results}
\end{table*}

\section{Evaluation}

Motivated by the results above, the remainder of our analysis focuses on the
\allconv{} model. To thoroughly investigate its performance and compare
it to the state of the art, we evaluate it on the following unseen datasets:
\begin{description}
\item[KeyFinder:]
    \num{1000} songs from a variety of popular music genres\footnote{\url{http://www.ibrahimshaath.co.uk/keyfinder/}}. Unfortunately,
    we have only the audio for \num{998} of the songs available.
\item[Isophonics:]
    \num{180} songs by The Beatles, \num{19} songs by Queen, and \num{18} songs
    by Zweieck\footnote{\url{http://isophonics.net/datasets}}. Since these songs contain key modulations, we split them into
    single key segments and retain only segments annotated as major or minor
    keys, as was done for the 2017 MIREX evaluation
    campaign\footnote{\url{http://www.music-ir.org/mirex/wiki/2017:Audio_Key_Detection_Results}}.
\item[Robbie Williams:]
    \num{65} songs by Robbie Williams, which we also split into single key
    segments as outlined above~\cite{digiorgi_automatic_2013}.
\item[Rock:]
    \num{200} songs taken from Rolling Stone's ``500 Greatest Songs of All
    Time'' list\footnote{\url{http://rockcorpus.midside.com/}}~\cite{declercq_corpus_2011}. As with the McGill Billboard dataset, only the tonics are
    annotated. We first split the songs according to the annotated tonics, and
    then follow a similar procedure as described
    in~\cite{korzeniowski_endtoend_2017}: if more than \SI{80}{\percent} of the
    tonic chords are in either major or minor, the mode is set accordingly; if
    there are no tonic chords in a segment, we consider dominant chords in the
    same way.
\end{description}

We select the best \allconv{} model based on the validation score over the compound data of Electronic, Pop/Rock and Classical music. On average, models with $\nf=20$ and dropout probability of
$0.1$ performed best. However, the best single model used $\nf=24$ (see
Fig.~\ref{fig:gridsearch}), and was consequently chosen as final model.

In Table~\ref{tab:results}, we compare this model to other models proposed in
the academic literature. For each dataset, we show the results of the best
competing system, if available. For the GiantSteps and Billboard datasets, the
best competing systems were variants of the
neural-network-based model from~\cite{korzeniowski_endtoend_2017}. For the
pre-segmented Isophonics and Robbie Williams datasets, we use the results
available on the MIREX 2017 website. For
the KeyFinder dataset, we report the best results achieved using the
open-source reference
implementation\footnote{\url{https://github.com/angelfaraldo/edmkey}} of the
algorithms from~\cite{faraldo_multiprofile_2017}.

\begin{figure}[t!]
\centering
\includegraphics{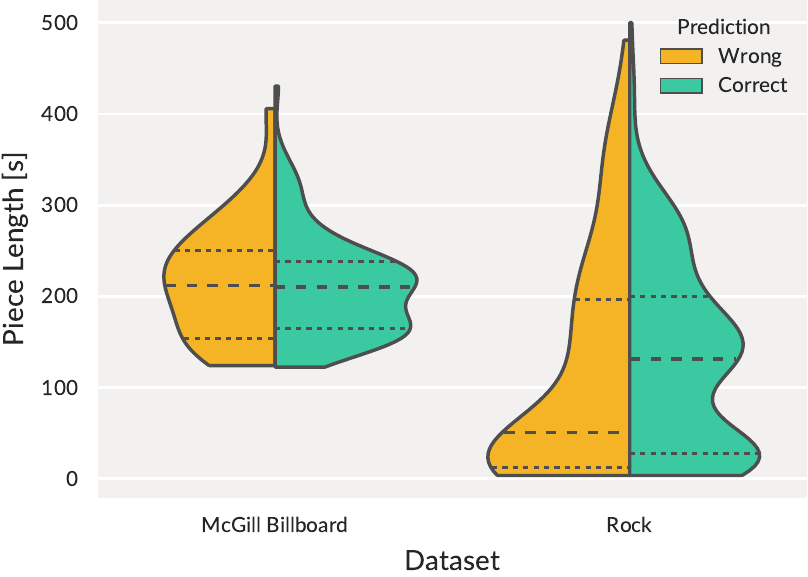}
\caption{Distributions of the length of correctly and incorrectly classified
        excerpts depending on the dataset they come from. Densities are
        estimated using kernel density estimation. Horizontal lines with long
        dashes indicate the median, those with short dashes the quartiles. The
        densities are normalised, i.e.\ they do not indicate how many instances
        were classified correctly (or incorrectly), but only the distribution
        of except lengths within each group.\vspace{-0.5em}}
\label{fig:duration_dependency}
\end{figure}

As we can see, the proposed model \emph{performs best for all datasets} for
which comparisons were possible. Keep in mind that the systems we compare to
are often specifically tuned for a genre (CK1,
CK2, HS1, bgate) or set up to favour certain key modes
prevalent in a dataset (BD1), while we use the same, general model for all
datasets. For example, CK1 performs badly on the Billboard
dataset ($w=72.8$), BD1 on the GiantSteps ($w=59.6$), and HS1 on the Isophonics
dataset ($w=64.1$). In this light, it is remarkable that the proposed model
consistently out-performs the others.

However, the results also point us to a deficiency of the model. Recall that
for some datasets (e.g.\ Rock), we split the files according to key
annotations, and process each excerpt individually. If we compare the
results on the Rock dataset with those on the Billboard dataset, we see a large
discrepancy, although both sets comprise similar musical styles. As
Fig.~\ref{fig:duration_dependency} demonstrates, the duration of a classified
excerpt plays a major role here: for the Billboard set, the median length of
excerpts classified correctly matches the one of incorrect classifications; for
the Rock set, however, the median lengths differ greatly: \SI{131}{\second}
vs.\ \SI{51}{\second}, for correctly and incorrectly classified excerpts,
respectively. The distribution of excerpt lengths that are classified
correctly is thus very different from the one of incorrectly classified excerpts in
the Rock set. The shorter an excerpt, the more likely it is classified
incorrectly.

This is not surprising per se. Determining the key of a piece requires a
certain amount of musical context. However, it shows that in order to move
beyond global key classification, and towards recognising key modulations, it
will not suffice to detect key boundaries and apply known methods within these
boundaries. To recognise key modulations, classifying short excerpts
individually will reach a glass ceiling. Instead, we will need models that
consider the hierarchical harmonic coherence of the whole piece.

\section{Conclusion}

We have presented a genre-agnostic key classification model based on the system
developed in~\cite{korzeniowski_endtoend_2017}, with improvements of the
training procedure and network structure. These improvements enable faster
training, better generalisation, and training larger and thus more powerful
models, which can leverage diverse training data instead of being
impaired by it. The resulting key classifier generalises well over datasets of
different musical styles, and out-performs systems that are specialised for
specific genres (see Table~\ref{tab:results}).

\section{Acknowledgements}

This work is supported by the European Research Council (ERC) under the EU's
Horizon 2020 Framework Programme (ERC Grant Agreement number 670035, project
``Con Espressione'').

\bibliography{ismir2018}

\end{document}